\documentstyle[aps,epsf,preprint]{revtex}

\begin{document}

\draft

\title{Muons and emissivities of neutrinos in neutron star cores}

\author{\O.\ Elgar\o y, L.\ Engvik and E.\ Osnes}

\address{Department of Physics, University of Oslo, N--0316 Oslo, Norway}

\author{F.\ V.\ De Blasio, M.\ Hjorth--Jensen and G.\ Lazzari}

\address{ECT*, European Centre for Theoretical Studies in
         Nuclear Physics and Related Areas, Trento, Italy}

\maketitle

\clearpage

\begin{abstract}
    In this work we consider the role of muons in various
    URCA processes relevant for neutrino emissions in the core
    region of neutron stars. The calculations are done for $\beta$--stable
    nuclear  matter with and without muons.
    We find muons
    to appear at densities $\rho = 0.15$ fm$^{-3}$,
    slightly around the saturation density
    for nuclear matter $\rho_0 =0.16$ fm$^{-3}$.  The direct
    URCA processes for nucleons are forbidden for densities below
    $\rho = 0.5$ fm$^{-3}$, however the modified URCA processes
    with muons  $(n+N\rightarrow p+N +\mu +\overline{\nu}_{\mu},
    p+N+\mu \rightarrow n+N+\nu_{\mu}$), where $N$ is a nucleon,
    result in neutrino emissivities comparable to those from
    $(n+N\rightarrow p+N +e +\overline{\nu}_e, p+N+e \rightarrow
    n+N+\nu_e$). This opens up for further possibilities to explain the
    rapid cooling of neutrons stars. Superconducting protons
    reduce however these
    emissivities at densities below $0.4$ fm$^{-3}$.
\end{abstract}

\pacs{PACS number(s): 97.60.Jd 21.65.+f 74.25.Bt }

\clearpage

The thermal evolution of a neutron star may provide information
about the interiors of the star, and in recent years much effort
has been devoted in measuring neutron star temperatures, especially
with the Einstein Observatory and ROSAT, see e.g.\ Ref.\ \cite{oegelman95}.
The main cooling mechanism in the early life of a neutron star is
believed to go through neutrino emissions in the core
of the neutron star \cite{pethick92,page94}.
The most powerful energy losses are expected to be given by the so--called
direct URCA mechanism
\begin{equation}
    n\rightarrow p +e +\overline{\nu}_e, \hspace{1cm} p+e \rightarrow
    n+\nu_e ,
    \label{eq:directU}
\end{equation}
as discussed recently by several authors \cite{pethick92,pplp92,prakash94}.
However, in the outer cores of massive neutron stars and in the
cores of not too massive neutron stars ($M < 1.3-1.4 M_{\odot}$), the direct
URCA process is allowed at densities
where the momentum conservation $k_F^n < k_F^p + k_F^e$ is
fulfilled\footnote{Throughout this work we will reserve letters $n$, $p$,
$e$ and $\mu$ to neutrons, protons, electrons and muons respectively. Also,
we will set $\hbar=c=1$.}. This happens
only at densities $\rho$ several times
the nuclear matter saturation density $\rho_0 =0.16$ fm$^{-3}$.

Thus, for long time the dominant processes for neutrino emission
have been the so--called modified URCA processes first discussed by
Chiu and Salpeter \cite{cs64}, in which the two reactions
\begin{equation}
    n+n\rightarrow p+n +e +\overline{\nu}_e,
    \hspace{0.5cm} p+n+e \rightarrow
    n+n+\nu_e ,
    \label{eq:ind_neutr}
\end{equation}
occur in equal numbers.
These reactions are just the usual processes of neutron
$\beta$--decay and electron capture on protons of Eq.\ (\ref{eq:directU}),
with the addition of an extra bystander neutron. They produce
neutrino--antineutrino pairs, but leave the composition of matter constant
on average. Eq.\ (\ref{eq:ind_neutr}) is referred to as the
neutron branch of the modified URCA process. Another branch is the
proton branch
\begin{equation}
    n+p\rightarrow p+p +e +\overline{\nu}_e, \hspace{0.5cm} p+p+e
    \rightarrow
    n+p+\nu_e ,
    \label{eq:ind_prot}
\end{equation}
pointed out by Itoh and Tsuneto \cite{it72} and recently reanalyzed
by Yakovlev and Levenfish \cite{yl95}. The latter authors showed that
this process is as efficient as Eq.\ (\ref{eq:ind_neutr}).
Similarly, at higher densities, if muons are present we have the
processes
\begin{equation}
    n\rightarrow p +\mu +\overline{\nu}_{\mu}, \hspace{1cm} p+\mu
    \rightarrow
    n+\nu_{\mu} ,
    \label{eq:directmu}
\end{equation}
\begin{equation}
  n+n\rightarrow n+p +\mu +\overline{\nu}_{\mu},
    \hspace {0.3cm} n+p+\mu \rightarrow n+n+\nu_{\mu},
    \label{eq:ind_neutmu}
\end{equation}
and
\begin{equation}
  n+p\rightarrow p+p +\mu +\overline{\nu}_{\mu},
    \hspace {0.3cm} p+p+\mu \rightarrow n+p+\nu_{\mu},
    \label{eq:ind_protmu}
\end{equation}
in addition one also has the possibility of neutrino--pair
bremsstrahlung,
processes with baryons more massive than the nucleon
participating, such as isobars or hyperons \cite{pplp92,prakash94},
or neutrino emission from more exotic states like pion and kaon
condensates  \cite{brown94,mstv90} or quark matter
\cite{glendenning91,iwamoto82,dth95}. Actually, Prakash {\em et al.}
\cite{pplp92} showed that the hyperon direct URCA processes
gave a considerable contribution to the emissivity,
without invoking exotic states or the large proton fractions needed
in Eq.\ (\ref{eq:directU}).

The scope of this Letter  is to give an estimate of the
processes of Eqs.\ (\ref{eq:directU})--(\ref{eq:ind_protmu})
at densities corresponding to the outer core of massive neutron stars
or the core of not too massive neutron stars, performing a self--consistent
Brueckner--Hartree--Fock (BHF)
calculation for $\beta$--stable nuclear matter. We
impose the relevant equilibrium conditions,
with and without muons, employing
modern meson--exchange potential models for the nucleon--nucleon
interaction.
In addition, we consider also the role of
superconducting protons in the core of the star, by calculating the pairing
gap for protons in the $^1S_0$  state.
The proton superconductivity  reduces considerably the energy losses
in the above reactions \cite{yl95},
and may have important consequences for the
cooling of young neutron stars.
The final outcome of our calculations yields then the equation
of state, effective
masses $m^*$ for protons and nucleons, their corresponding
Fermi momenta $k_F$, the Fermi momenta for electrons and muons
and the proton pairing gap. These ingredients enter into the
calculation of the various reaction rates.

In more detail, our calculational procedure is as follows:\newline
First we solve self--consistently the BHF
 equations
for the single-particle energies,
using a $G$--matrix defined through the Bethe--Brueckner--Goldstone
equation as
\begin{equation}
   G=V+V\frac{Q}{\omega -H_0}G,
\end{equation}
where $V$ is the nucleon-nucleon potential, $Q$ is the Pauli operator
which prevents scattering into intermediate
states prohibited by the Pauli
principle, $H_0$ is the unperturbed
hamiltonian acting on the intermediate
states and $\omega$ is the so--called starting energy,
the unperturbed energy
of the interacting states. Methods to solve this equation are reviewed in
Ref.\ \cite{hko95}.
The single--particle energies for state $k_i$ ($i$
encompasses all relevant
quantum numbers like momentum,
isospin projection, spin etc.)
in nuclear matter are assumed to have
the simple quadratic form
\begin{equation}
   \varepsilon_{k_i}=
   {\displaystyle\frac{k_{i}^2}
   {2m^{*}}}+\delta_i ,
   \label{eq:spen}
\end{equation}
where $m^{*}$ is the effective mass.
The terms $m^{*}$ and $\delta$, the latter being
an effective single--particle
potential related to the $G$--matrix, are obtained through the
self--consistent BHF  procedure. The so--called
model--space BHF method
for the single--particle spectrum has been used, see
e.g.\ Refs.\ \cite{hko95}, with a cutoff $k_M=3.0$ fm$^{-1}$.
This self--consistency scheme
consists in choosing adequate initial values of the
effective mass and $\delta$. The obtained $G$--matrix is in turn used to
obtain new values for $m^{*}$ and $\delta$ for protons and neutrons.
This procedure
continues until these parameters vary little.
The BHF equations are solved
for different proton fractions, using the formalism of Refs.\
\cite{hko95,ks93}. The conditions of $\beta$ equilibrium require
that
\begin{equation}
     \mu_n=\mu_p+\mu_e,
\end{equation}
where $\mu_i$ is the chemical potential
and that charge is conserved
\begin{equation}
     \rho_p=\rho_e,
\end{equation}
where $\rho_i$ is the particle number density in fm$^{-3}$ for particle $i$.
If muons are present, we need also to satisfy charge conservation
\begin{equation}
     \rho_p=\rho_e+\rho_{\mu},
\end{equation}
and energy conservation through
\begin{equation}
     \mu_e=\mu_{\mu}.
\end{equation}
The nucleon--nucleon potential is defined by the
parameters of the meson--exchange potential model of the Bonn group,
version A in Table A.2 of Ref.\ \cite{mac89}.
{}From the derived effective  interaction one can in turn
calculate the energy per nucleon as function of the total
density \cite{hko95}. This is plotted in Fig.\ \ref{fig:fig1}
with and without muons.
As can be seen from this figure, muons yield contributions
to the energy per particle (and in Fig.\ \ref{fig:fig2} to the pairing
gap)
at densities around $0.15$ fm$^{-3}$.
The presence of muons makes the energy per particle  and in turn
the equation of state softer, yielding smaller masses and larger radii
($M=1.45 M_{\odot}$ and $R\approx 9$ km here)  for neutron stars.
Similar qualitative results are reached if one performs
a Dirac--Brueckner--Hartree-Fock calculation,
as in Ref.\ \cite{ehobo94},
although the equation of state in that case is stiffer, resulting
in larger masses and smaller radii for the star.
More details and applications to neutron stars
will be reported by us in a future work \cite{eeodhl95}.

The next step is to evaluate the gap equation following the scheme
proposed by Anderson and Morel \cite{am61} and applied
to nuclear physics
by Baldo {\em et al.\ } \cite{bcll90}. These authors introduced an
effective interaction $\tilde{V}_{k,k'}$.
This effective interaction
sums up all two--particle excitations
above the cutoff $k_M$. It is defined
according to
\begin{equation}
     \tilde{V}_{k,k'}=V_{k,k'}-\sum_{k''>k_M}V_{k,k''}\frac{1}{2E_{k''}}
                      \tilde{V}_{k'',k'},
     \label{eq:gap1}
\end{equation}
where the quasiparticle energy $E_k$ is given by
\begin{equation}
     E_k=\sqrt{\left(\varepsilon_k-\varepsilon_F\right)^2+\Delta_k^2},
     \label{eq:gap2}
\end{equation}
$\varepsilon_F$ being the single--particle energy at the Fermi surface,
$V_{k,k'}$ is the free nucleon--nucleon potential in momentum space
and $\Delta_k$ is the pairing gap
\begin{equation}
  \Delta_k=-\sum_{k'\leq k_M}\tilde{V}_{k,k'}\frac{\Delta_{k'}}{2E_{k'}}.
     \label{eq:gap3}
\end{equation}
For notational economy, we
have dropped the subscript $i$ on the single--particle energies.

In summary, first we obtain the self--consistent
BHF single--particle spectrum $\varepsilon_k$ for protons and neutrons
in $\beta$--stable matter. This procedure gives the relevant
proton and neutron fractions at a given total density.
Thereafter we solve self--consistently Eqs.\ (\ref{eq:gap1}) and
(\ref{eq:gap3}) in order to obtain the pairing gap
$\Delta$ for protons or neutrons.
For states above $k_M$, the quasiparticle energy of
(\ref{eq:gap2}) is approximated by
$E_k=\left(\varepsilon_k-\varepsilon_F\right)$, an
approximation found to yield satisfactory results in neutron matter
\cite{elgaroey95}.
Our prescription gives then the pairing gap at $T=0$. The results
for the proton pairing gap with and without muons are displayed in
Fig.\ \ref{fig:fig2} as functions of the total baryonic density.

The reader
should notice that our calculation of the pairing gap does
not include complicated many--body contributions  like polarization
terms. These may further reduce the pairing gap
and bring in an uncertainty regarding the size, which
strongly influences the modified URCA processes.
Although we have employed the Bonn A potential in Table A.2
of Ref.\ \cite{mac89}, the pairing gap obtained with other potentials
like the Paris or Bonn B and Bonn C is rather similar.
results. This is expected since the $^1S_0$ channel is determined
by the central force component only of the nucleon--nucleon
interaction, and all modern potentials  reproduce
the phase shifts for this channel.

With the pairing gap, effective masses and the Fermi momenta
for the various particles, we first look at
the neutrino energy production rate for Eq.\ (\ref{eq:ind_neutr}).
The expressions for these
processes
were derived by Friman and Maxwell \cite{fm79} and read \cite{yl95}
\begin{eqnarray}
        Q_{n}& \approx 8.5\times 10^{21}
        \left(\frac{m_n^*}{m_n}\right)^3
        \left(\frac{m_p^*}{m_p}\right)
        \left(\frac{\rho_e}{\rho_0}\right)^{1/3}
        T_9^8\alpha_n\beta_n,
        \label{eq:neutron_branch}
\end{eqnarray}
in units of ergs cm$^{-3}$s$^{-1}$ where
$T_9$ is the temperature in units of $10^9$ K, and according
to Friman and Maxwell $\alpha_n$ describes the momentum transfer
dependence of the squared matrix element in the Born approximation
for the  production rate in the neutron branch. Similarly,
$\beta_n$ includes the non--Born corrections and corrections due
to the nucleon--nucleon interaction not described by one--pion
exchange. Friman and Maxwell \cite{fm79} used $\alpha_n \approx 1.13$
at nuclear matter saturation density
and $\beta_n=0.68$. In the results presented below, we will not
include $\alpha_n$ and $\beta_n$. For the reaction of
Eq.\ (\ref{eq:ind_neutmu}), one has to replace $\rho_e$ with $\rho_{\mu}$.
For the proton branch of Eq.\ (\ref{eq:ind_prot})
we have the approximate equation \cite{yl95}
\begin{eqnarray}
        Q_{p}&\approx 8.5\times 10^{21}
        \left(\frac{m_p^*}{m_p}\right)^3
        \left(\frac{m_n^*}{m_n}\right)
        \left(\frac{\rho_e}{\rho_0}\right)^{1/3}\nonumber \\
        &\times T_9^8\alpha_p\beta_p
        \left(1-\frac{k_F^e}{4k_F^p}\right)\Theta ,
        \label{eq:proton_branch}
\end{eqnarray}
where $\Theta =1$ if $k_F^n < 3k_F^p+k_F^e$ and zero
elsewhere. Yakovlev and Levenfish
put $\alpha_n=\alpha_p$ and $\beta_p=\beta_n$. We will, due to the
uncertainty in the determination of these coefficients, omit them
in our calculations of the reaction rates. For Eq.\ (\ref{eq:ind_protmu})
one replaces $\rho_e$ with $\rho_{\mu}$ and
$k_F^e$ with $k_F^{\mu}$.

The reaction rates for the
URCA processes are reduced due to the superconducting
protons. Here we adopt the results from Yakovlev and Levenfish
\cite{yl95}, their Eqs.\ (31) and (32) for the neutron
branch of Eq.\ (\ref{eq:neutron_branch}) and Eqs.\ (35) and (37)
for the proton branch of Eq.\ (\ref{eq:proton_branch}).
In this work we study proton singlet--superconductivity  only,
employing the standard approximation \cite{aoe85}
\begin{equation}
       \frac{k_BT_C}{\Delta (0)}=0.5669,
\end{equation}
where $T_C$ is the critical temperature
and $\Delta (0)$ is the pairing gap at zero temperature derived in
our calculations. The critical temperature is then used to
obtain the temperature dependence of the corrections
to the neutrino reaction rates due to superconducting
protons. To achieve that we employ Eq.\ (23) of \cite{yl95}.

In this Letter, we present results for the neutrino energy
rates at a density $\rho =0.3$ fm$^{-3}$.
The critical temperature is $T_C=3.993\times 10^{9}$ K, whereas without
muons we have $T_C=4.375\times 10^{9}$.
The implications for the final neutrino rates are shown in
Fig.\ \ref{fig:fig3}, where we show the results for the full case
with both muons and electrons for the
processes of Eqs.\ (\ref{eq:ind_neutr}),
(\ref{eq:ind_prot}), (\ref{eq:ind_neutmu}) and (\ref{eq:ind_protmu}).
In addition, we also display the results when there is no reduction
due to superconducting protons.
At the density considered, $\rho=0.3$ fm$^{-3}$,
we see that the processes of Eqs.\
(\ref{eq:ind_neutmu}) and (\ref{eq:ind_protmu}) are comparable
in size to those of Eqs.\
(\ref{eq:ind_neutr}) and (\ref{eq:ind_prot}) (on the log--scale
there is basically no difference).
The direct URCA process of Eq.\ (\ref{eq:directU}) is
not allowed in our $\beta$--stable matter
for densities
$\rho<0.5$ fm$^{-3}$, due to momentum
conservation. Similarly, the direct URCA process
with muons is also not allowed
at densities $\rho <0.6$ fm$^{-3}$. Thus, Eqs.\ (\ref{eq:ind_neutmu})
and (\ref{eq:ind_protmu}) give an additional and important
contribution to the neutrino production in the core of a neutron
star.
However,
the proton pairing gap is still sizable at densities up to
$0.4$ fm$^{-3}$, and yields a significant suppression of the
modified URCA processes discussed here, as seen in Fig.\ \ref{fig:fig3}.
We have omitted any discussion on neutron pairing in the $^3P_2$ state.
For this channel we find the pairing gap to be
rather small \cite{elgaroey95}, less
than $0.1$ MeV and close to that obtained in Ref.\ \cite{tt93}.
We expect therefore that the  major reductions of the neutrino rates
in the core come from superconducting protons in the $^1S_0$ state.
Moreover, the recent reinvestigation  of neutrino--pair bremsstrahlung
in the crust by Pethick and Thorsson \cite{pt94} indicates that this process
is much less important for the thermal evolution of neutrons stars than
suggested by
earlier calculations. Thus, combined with the present reduction
due to superconducting protons, our results may indicate that there
is need for other processes than those studied here to explain
the rapid cooling of neutron stars. The
contribution from neutrino--pair
bremsstrahlung in nucleon--nucleon collisions in the core
is also, for most temperature
ranges relevant for neutron stars,
smaller  than the contribution from modified URCA processes \cite{yl95}.
Possible candidates are then direct URCA processes due to hyperons
and isobars, as suggested in Ref.\ \cite{pplp92}, or neutrino production
through exotic states of matter, like kaon or pion condensation
\cite{brown94,mstv90} or quark matter
\cite{glendenning91,iwamoto82,dth95}. Though, the analysis
of Page \cite{page94}
indicates that it is hard to discriminate
between fast or slow cooling scenarios, though in both cases
agreement with the observed temperature of Geminga is
obtained if baryon pairing is present in most, if not all
of the core of the star.

In summary, we have performed a self--consistent calculation
for $\beta$--stable neutron matter, with and without muons.
We have shown that the modified URCA processes
$(n+N\rightarrow p+N +\mu +\overline{\nu}_{\mu},
p+N+\mu \rightarrow n+N+\nu_{\mu}$), where $N$ is a nucleon,
result in neutrino emissivities comparable to those from
$(n+N\rightarrow p+N +e +\overline{\nu}_e, p+N+e \rightarrow
n+N+\nu_e$). These processes should therefore be accounted
for in a scenario for neutron star cooling.

This work has been supported by the Istituto Trentino di Cultura, Italy,
the Research Council of Norway and the NorFA (Nordic Academy for
Advanced Research).

\begin{figure}[hbtp]
     \setlength{\unitlength}{1mm}
     \begin{picture}(100,70)
     \end{picture}
     \caption{Energy per baryon as function of the total
              baryonic density $\rho$ in $\beta$--stable neutron matter
              with (dashed line) and without (solid line) muons.}
     \label{fig:fig1}
\end{figure}

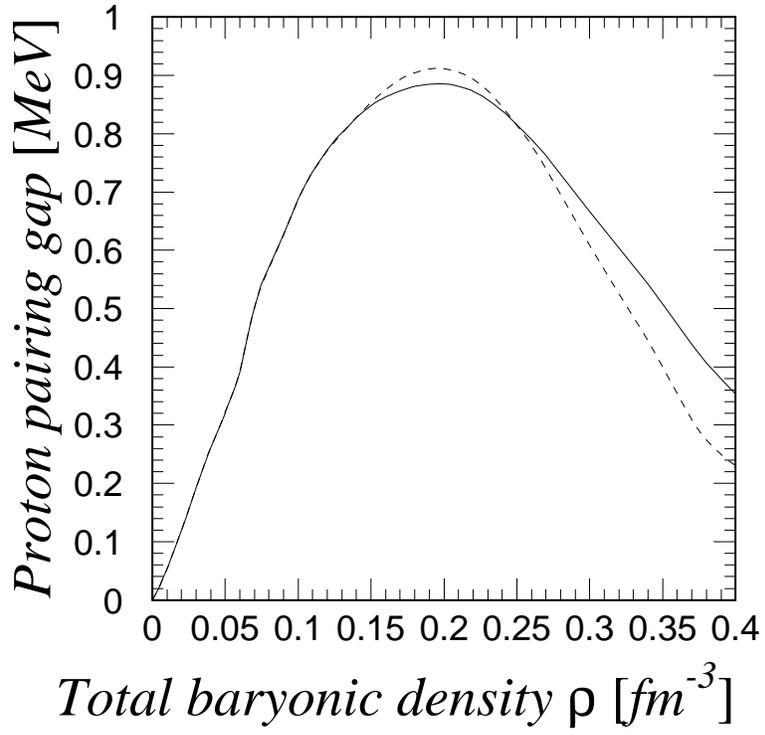
\begin{figure}[hbtp]
     \setlength{\unitlength}{1mm}
     \begin{picture}(100,70)
     \end{picture}
     \caption{Proton pairing energy gap as function of total
              baryonic density $\rho$ in $\beta$--stable neutron matter
              with (dashed line) and without (solid line) muons.}
     \label{fig:fig2}
\end{figure}

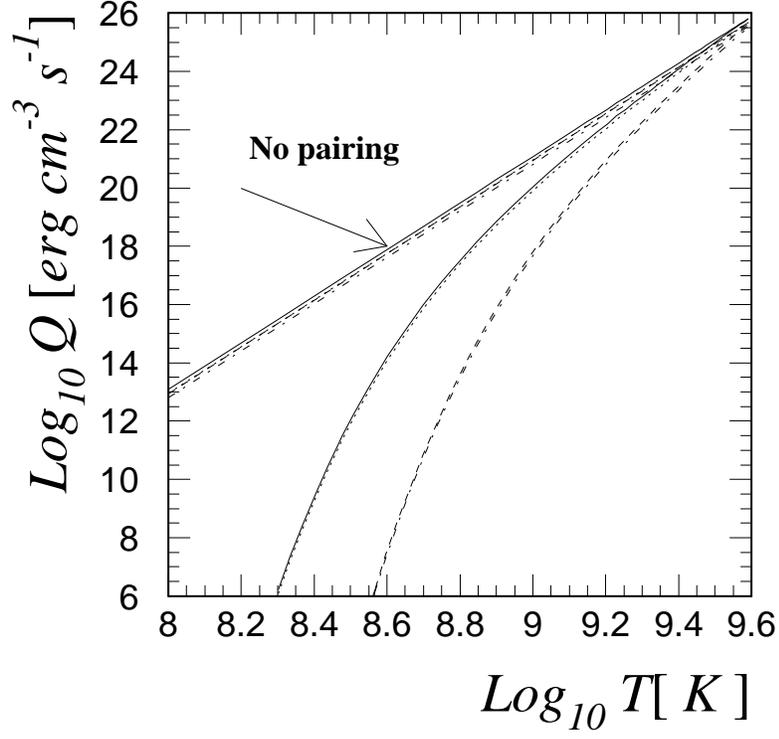
\begin{figure}[hbtp]
     \setlength{\unitlength}{1mm}
     \begin{picture}(100,70)
     \end{picture}
     \caption{Temperature dependence of neutrino energy loss
              rates in a neutron star core at a total baryonic
              density of $0.3$ fm$^{-3}$ with muons and electrons.
              Solid line represents Eq.\ (2), dotted line is Eq.\ (5),
              dashed line is Eq.\ (3) while the dash--dotted line
              is the processes of Eq.\ (6). The corresponding results
              with no pairing are also shown.}
     \label{fig:fig3}
\end{figure}

\end{document}